\documentclass{KapProc} 
\kluwerbib
\let\lcitebracket(
\let\rcitebracket)
\usepackage[dvips]{graphics,epsfig}
\newcommand{\lsim}{\raise0.3ex\hbox{$<$}\kern-0.75em{\lower0.65ex\hbox{$\sim$}}}
\newcommand{\gsim}{\raise0.3ex\hbox{$>$}\kern-0.75em{\lower0.65ex\hbox{$\sim$}}}
\newcommand{\Msolar}{\mbox{\,$\rm M_{\odot}$}}        
\newcommand{\Mblack}{\mbox{\,$\rm M_{bh}$}}     
\newcommand{\Mbulge}{\mbox{\,$\rm M_{bulge}$}} 
\newcommand{\Lbulge}{\mbox{\,$\rm L_{bulge}$}} 
\newcommand{\Ltot}{\mbox{\,$\rm L_{tot}$}} 
\newcommand{\sbsc}[2]{\mbox{$\rm #1_{#2}$}}     

\begin{document}


\articletitle{The radio loudness dichotomy: \\
environment or black-hole mass?}
\author{Ross McLure$^1$, James Dunlop$^2$} 
\affil{$^1$University of Oxford, UK\\
$^2$Institute for Astronomy, University of Edinburgh, UK\\}
\begin{abstract}
\noindent
The results of a comprehensive study of the cluster environments and 
black-hole masses of an optically matched sample of radio-loud and 
radio-quiet quasars are presented. No evidence is found for a difference 
in large-scale environments, with both quasar classes found to be 
located in clusters of Abell class $\sim0$ . Conversely, virial 
black-hole mass estimates based on H$_{\beta}$
line-widths show a clear difference in the quasar black-hole mass 
distributions. Our results suggest that a black-hole mass of
$\sim10^{9}\Msolar$ is required to produce a powerful radio-loud
quasar, and that it is black-hole mass and accretion rate which hold 
the key to the radio-loudness dichotomy. 
\end{abstract}

\section*{Cluster environments of powerful AGN at $z\simeq0.2$}

The sample for this study consists of 44 powerful AGN with redshifts
in the range $0.1<z<0.3$. The full sample is comprised of three matched
sub-samples of 10 radio galaxies (RGs), 13 radio-loud quasars (RLQs)
and 21 radio-quiet quasars (RQQs), the majority of which are drawn
from the {\sc hst} host-galaxy study of McLure et al.\ (1999) and
Dunlop et al.\ (2001), with further objects taken from the 
host-galaxy study of Bahcall et al.\ (1997). The sub-samples are 
matched such that 
the $L_{5GHz}-z$ distribution of the two radio-loud sub-samples, 
and the $M_{V}-z$ distribution of the two quasar sub-samples are 
statistically indistinguishable.

Counts of excess galaxies surrounding each AGN were used to 
calculate the spatial clustering amplitude ($B_{gq}$) in order 
to provide a quantitative estimate of the richness of their cluster 
environments \cite{ls79}. A full description of the analysis and results 
of this study can be found in McLure \& Dunlop (2001a).

\subsection*{Do RLQs have richer environments than RQQs?}

\begin{figure}[ht]
\centerline{\epsfig{file=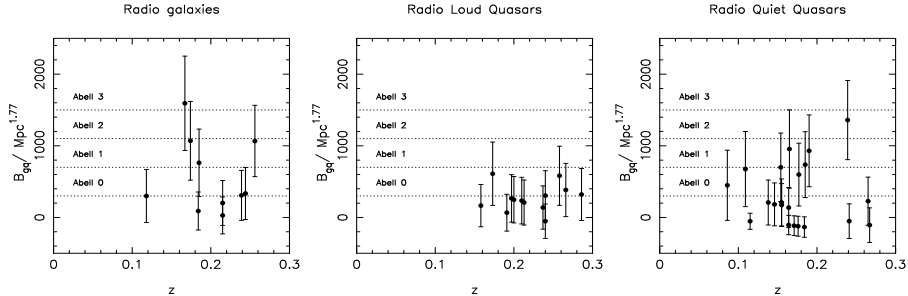,width=12cm,angle=0,clip=}}
\caption{The B$_{gq}-z$ distribution for the three sub-samples.  Also
shown are approximate Abell cluster classifications according to the 
linear scheme proposed by Yee \& L{\'o}pez-Cruz (1999), renormalized 
such that $B_{gq}=300$Mpc$^{1.77}$ corresponds to Abell class 0.}
\label{fig1}
\end{figure}
The spatial clustering amplitude results for the three sub-samples
are shown in Fig \ref{fig1}. It can be seen that there is no
indication that the RLQs occupy systematically richer
environments than the RQQs, with their respective mean clustering 
amplitudes of $267\pm51$ and $326\pm94$Mpc$^{1.77}$ indicating
that both quasar classes occupy cluster environments comparable to
Abell class 0. We note here that our finding that RQQs occupy similar
environments to RLQs is not in general agreement with the literature,
with the majority of previous studies finding RQQs to inhabit poorer
environments than RLQs. For example, both Smith, Boyle \& Maddox
(1995) and Ellingson, Yee \& Green (1991) found that at $z<0.3$ RQQs
have environments perfectly
consistent with those of field galaxies. Although the exact cause of this
discrepancy is difficult to determine, it is probably due to the fact
that the RQQs and RLQs in our study our well matched in terms of 
nuclear luminosity.

\begin{figure}[ht]
\centerline{\epsfig{file=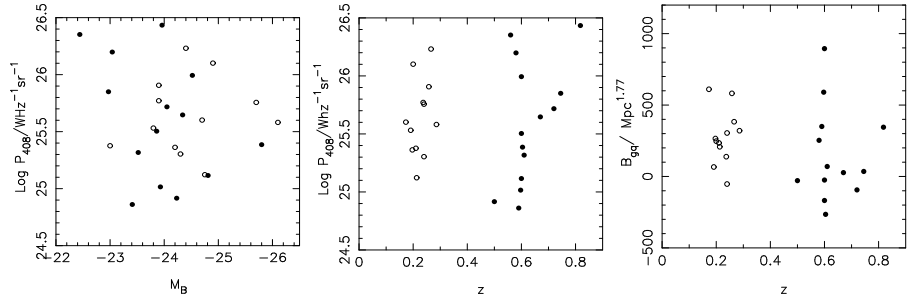,width=12.0cm,angle=0,clip=}}
\caption{The left-hand and middle panels show the matching between our
RLQ sample (open circles) and the sub-sample of the objects studied by
Wold et al.\ (2000) discussed in the text (filled circles). The right-hand
panel shows a comparison of their respective spatial clustering amplitudes.}
\label{fig2}
\end{figure}

Further support for this argument is provided by the good agreement
between our results and those of the recent study by Wold et
al.\ (2000; 2001; these proceedings) which involved a 
similar analysis of an
optically matched sample of RQQs and RLQs at redshifts in the range
$0.5<z<0.8$. Wold et al. also conclude that there is no difference in
the average cluster richness of RQQs and RLQs of similar optical
luminosity, suggesting that our results persist at higher redshift. In
Fig \ref{fig2} we show a comparison of the clustering results for
our RLQ sample, and those for a sub-sample of the Wold et al. RLQs
which display the same range in radio luminosity. It is clear from Fig
\ref{fig2} that there is no evidence for evolution in 
RLQ cluster environments, at least out to $z\sim 0.8$.

The lack of any significant difference in the
cluster environments of optically-matched samples of RQQs and RLQs
suggests that cluster environment cannot be the primary cause of the
quasar radio-loudness dichotomy. In the following section we move on
from studying the large-scale environments of quasars, to explore
what role (if any) is played by black-hole mass in determining the 
radio luminosity of AGN.   

\section*{The black-hole masses of quasars and Seyfert galaxies}

In this study \cite{md01b} the quasar sample from
 host-galaxy study of McLure et al.\ (1999)
and Dunlop et al.\ (2001) was combined with a sample of 15
Seyfert 1 galaxies from the reverberation mapping study of Wandel,
 Peterson \& Malkan (1999). This combined sample provided the 
opportunity to 
investigate the relation between black-hole mass ($\Mblack$) 
and host-galaxy bulge mass ($\Mbulge$) over a wide range in AGN 
nuclear luminosity, and was assembled with the aim of addressing 
two important questions. Firstly, is the ratio of $\Mblack/\Mbulge$ 
in Seyfert galaxies really a factor
of twenty lower than seen in nearby galaxies and quasar hosts, 
as claimed by Wandel (1999), or can the apparent discrepancy be 
explained by a systematic over-estimate of the Seyfert bulge luminosities? 
Secondly, do differences in black-hole masses and gas accretion 
rates play a crucial role in determining the radio properties of AGN?

\begin{figure}[hb]
\centerline{\epsfig{file=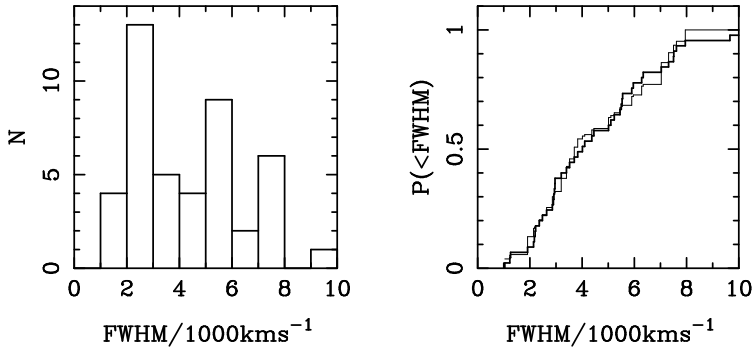,width=10.0cm,angle=0,clip=}}
\caption{The left-hand panel shows the distribution of
$\sbsc{H}{\beta}$ {\sc fwhm} measurements for the 45 objects in the
combined quasar+seyfert sample. The right-hand panel shows the 
cumulative {\sc fwhm}
distributions displayed by the data (thick line) and that of the disc
BLR model discussed in the text (thin line). The two cumulative 
distributions are indistinguishable, with a KS test probability of $p=0.99$}
\label{fig3}
\end{figure}

Host-galaxy bulge luminosities for the vast majority (39/45) of the
sample were determined via full two
dimensional disc/bulge decomposition of high resolution {\sc hst}
data, with the luminosities of the remaining objects being taken 
from literature disc/bulge decompositions. Consequently, our revised 
Seyfert bulge luminosities should be more accurate than
the $\Lbulge/\Ltot$ morphology-based estimates adopted by Wandel
(1999). The black-hole masses are virial estimates \cite{wpm99}
 where it is assumed that the velocities of the
broad-line regions clouds (as estimated from the {\sc fwhm} of the
H$_{\beta}$ emission line) are due to the gravitational potential of
the central black-hole. We make the further assumption that the
broad-line region has a flattened disc-like geometry, and that the
H$_{\beta}$ line-widths are therefore orientation dependent, as
suggested by the results of Wills \& Browne (1986). The 
success of this model in 
reproducing the observed line-width distribution is 
illustrated in Fig \ref{fig3}.

\section*{The $\Mblack-\Mbulge$ relation}
\begin{figure}[ht]
\centerline{\epsfig{file=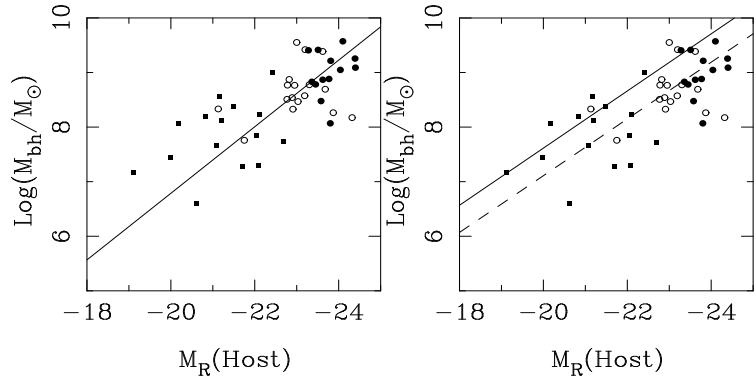,width=10.0cm,angle=0,clip=}}
\caption{Both panels show black-hole mass vs. host galaxy $R$-band
magnitude. The quasars are shown as open (radio quiet) and filled
(radio loud) circles, while the Seyfert galaxies are shown as filled
squares. In the left-hand panel the solid
line is the best fit to the data. The solid line in the right-hand
panel is the predicted relation from Magorrian et al.\ (1998). The 
dashed line in the right-hand panel is best fit to the data forcing a
 constant $\Mblack/\Mbulge$ ratio, and corresponds 
to $\Mblack/\Mbulge=0.0025$.}  
\label{fig4}
\end{figure}
The black-hole mass vs. bulge luminosity relation for the combined
quasar and Seyfert galaxy sample is plotted in Fig \ref{fig4}. The
two quantities can be seen to be well correlated, with the
rank-order coefficient of $\sbsc{r}{s}=-0.66$ having a significance of
$4.4\sigma$. It can be seen immediately that, with our revised bulge
luminosity estimates, there is no evidence that the Seyfert galaxies
follow a different relation to the more luminous quasars. 
The best-fitting relation ($\chi^{2}=64.8$ for 45 d.o.f.) is found to be:
\begin{equation}
\log(\Mblack/\Msolar)=-0.61(\pm0.08){\rm M}_{\rm R}-5.41(\pm1.75)
\end{equation}
\noindent
and is shown as the solid line in the left-hand panel of Fig
\ref{fig4}. It is noteworthy that the slope expected from a 
linear $\Mblack-\Mbulge$ relation
is $-0.52$, which is not inconsistent with the best-fitting value of 
$-0.61\pm0.08$. Consequently, a least-squares fit of
the data was undertaken with an enforced slope of $-0.52$, and is shown
as the dashed line in the right-hand panel of Fig \ref{fig4}. This 
can be seen to be a reasonable representation of the data, and
corresponds to a relationship of the form
$\Mblack=0.0025\Mbulge$. For reference, the solid line in right-hand 
panel of Fig \ref{fig4} shows the predicted relation from Magorrian et
al.\ (1998), 
which can be seen to systematically over predict the 
black-hole masses by a factor of $\sim2.5$.

The black-hole mass estimates for the 17 RQQs and 13 RLQs 
quasars provide an opportunity to determine the
influence (if any) of black-hole mass on quasar radio luminosity. The
continuum luminosity distributions of the two quasar
sub-samples are indistinguishable, implying that any difference in
their black-hole mass distributions are presumably linked to the 
difference in radio properties. 

Our results suggests that a difference does exist between the RQQ and
RLQ black-hole mass distributions, with the median black-hole mass of the
RLQs being a factor of three larger than their radio-quiet
counterparts. A natural
division between the quasar sub-samples appears to occur at
$\Mblack\sim10^{8.8}\Msolar$. Only 2/13 of the radio-loud quasars have
$\Mblack<10^{8.8}\Msolar$, while only 4/17 of the radio-quiet have
$\Mblack>10^{8.8}\Msolar$. This difference in black-hole mass 
distributions is shown to be significant at the $2.9\sigma$ ($p=0.004$) 
level by a KS test. The implication from the
quasar sample is therefore that (albeit with substantial overlap) 
for a given nuclear luminosity, the
probability of a source being radio-loud increases with black-hole
mass, or alternately decreases with ${\rm L}/\sbsc{L}{Edd}$. The same 
conclusion was recently reached by Laor (2000) from his study of 
the virial black-hole masses of PG quasars.

\section*{Conclusions}
\begin{itemize}
\item{We find no detectable difference in the average cluster
environments of optically matched RQQs and RLQs at $z\sim 0.2$,
implying that cluster richness is not the primary cause of the
radio-loudness dichotomy. } 
\item{Our results suggest that on average, luminous quasars occupy cluster
environments as rich as Abell class 0, although a large scatter is present.}
\item{We find that the bulges of both Seyfert galaxies and quasar host
galaxies follow the same relation between black-hole and bulge mass,
with a best-fitting linear relation of $\Mblack=0.0025\Mbulge$.} 
\item{Our black-hole mass estimates suggest that black-hole mass is
a key parameter in determining an AGN's radio luminosity, and that a
black-hole of mass $\simeq 10^{9}\Msolar$ is required to produce a
powerful radio-loud quasar.}
\end{itemize}



\begin{chapthebibliography}{<widest bib entry>}

\bibitem[Bahcall et al. 1997]{bah97}
Bahcall J.N., Kirhakos S., Saxe D.H., Schneider D.P., 1997, ApJ, 479, 642

\bibitem[Dunlop et al. 2001]{jsd01}
Dunlop J.S., et al., 2001, MNRAS, submitted

\bibitem[Ellingson, Yee \& Green 1991]{eyg91}
Ellingson E., Yee H.K.C., Green R.F., 1991, ApJ, 371, 49

\bibitem[Longair \& Seldner 1979]{ls79}
Longair M.S., Seldner M., 1979, MNRAS, 189, 433

\bibitem[Laor 2000]{l2000}
Laor A., 2000, ApJ, 543, L111

\bibitem[Magorrian et al. 1998]{mag98}
Magorrian J., et al., 1998, AJ, 115, 2285

\bibitem[McLure \& Dunlop 2001a]{md01}
McLure R.J., Dunlop J.S., 2001, MNRAS, in press (astro-ph/0007219)

\bibitem[McLure \& Dunlop 2001b]{md01b}
McLure R.J., Dunlop J.S., 2001, MNRAS, submitted (astro-ph/0009406)

\bibitem[McLure et al.\ 1999]{m99}
McLure R.J., Kukula M.J., Dunlop J.S., Baum S.A., O'Dea C.P., Hughes D.H.,
1999, MNRAS, 308, 377

\bibitem[Smith, Boyle \& Maddox 1995]{smith95} 
Smith R.J., Boyle B.J., Maddox S.J., 1995, MNRAS, 277, 270

\bibitem[Wandel 1999]{w99}
Wandel A., 1999, ApJ, 519, L39

\bibitem[Wandel, Peterson \& Malkan 1999]{wpm99} 
Wandel A., Peterson B.M., Malkan M.A., 1999, ApJ, 526,579

\bibitem[Wills \& Browne 1986]{wb86}
Wills B.J., Browne I.W.A., 1986, ApJ, 302, 56

\bibitem[Wold et al.\ 2000]{wold00}
Wold M., Lacy M., Lilje P.B., Serjeant S., 2000, MNRAS 316,267

\bibitem[Wold et al.\ 2001]{wold01}
Wold M., Lacy M., Lilje P.B., Serjeant S., 2001, MNRAS, 
in press (astro-ph/0011394)

\bibitem[Yee \& L{\'o}pez-Cruz 1999]{ylc99}
Yee H.K.C., L{\'o}pez-Cruz O., 1999, AJ, 117, 1985

\end{chapthebibliography}
\end{document}